\documentclass[conference,usenatbib]{basi}
\usepackage[T1]{fontenc}
\usepackage[british]{babel}
\usepackage[varg]{txfonts}
\usepackage{rotating,dcolumn,graphicx}

\begin{document}
\title[Double-Peaked Emission Line Seyferts]{The Search for Binary Black holes in Seyferts with Double Peaked Emission Lines}
\author[P. Kharb et~al.]
       {P.~Kharb$^1$\thanks{email: \texttt{kharb@iiap.res.in}},
       M.~Das$^{1}$, S.~Subramanian$^1$ and Z.~Paragi$^2$\\
       $^1$Indian Institute of Astrophysics, Koramangala II Block, Bangalore 560034, India\\
       $^2$Joint Institute for VLBI ERIC, Postbus 2, 7990 AA Dwingeloo, the Netherlands}
\pubyear{2015}
\volume{}
\pagerange{\pageref{firstpage}--\pageref{lastpage}}
\date{Received --- ; accepted ---}
\maketitle
\label{firstpage}

\begin{abstract}
We discuss results from very long baseline interferometry (VLBI) observations of two Seyfert galaxies with double peaked emission lines in their SDSS optical spectra. Such AGN are potential candidates for the presence of binary black holes, which can be resolved on parsec-scales with VLBI. Our observations do not detect twin radio cores but rather nuclear outflows in these Seyferts. These outflows could be interacting with the emission line clouds producing the double peaks in the emission lines.
\end{abstract}
\begin{keywords}
galaxies: nuclei -- galaxies: Seyfert -- radio continuum: galaxies
\end{keywords}

\section{Introduction}
Carl Seyfert noted in 1943, the existence of a small fraction of spiral galaxies that had bright star-like nuclei and peculiar emission line spectra populated by broad and narrow high-ionisation emission lines. We now know that these Seyfert galaxies are a subset of the class of sources hosting active galactic nuclei (AGN). The extreme luminosities of AGN, which can exceed the total light from the host galaxy by a factor of ten to a thousand, are understood to be a consequence of the release of gravitational energy as matter gets accreted onto a supermassive black hole (mass $\sim10^6-10^9$\,M$_\odot$). The emission lines are produced in dense fast-moving gas clouds (velocities $\sim100-10,000$\,km\,s$^{-1}$) that orbit around the black hole-accretion disk system, inside what are known as the broad and narrow line regions (BLR and NLR). 
A small fraction of AGN exhibit double peaks in their emission line spectra (e.g., Figure~1), indicative of a different kind of activity in their central regions.

Three popular scenarios have been put forward to explain double peaked emission line AGN (DPAGN): (i) their emission line regions are in rotating disks, (ii) their emission line clouds are being pushed out by radio outflows originating in the central engine, and (iii) there are binary black holes in the AGN carrying their respective emission line regions with them \citep[e.g.,][]{Gaskell83,Rosario10,Smith12,Gabanyi14}. However, binary black holes are expected to be rare in spiral galaxies that typically host Seyfert nuclei. The hierarchical galaxy formation model \citep[e.g.,][]{Steinmetz02} proposes that spiral galaxies merge to form elliptical galaxies, while they themselves only grow through minor mergers which may or may not result in binary black holes \citep{Aguerri01}. $<$1\% of DPAGN are likely to have binary black holes with kiloparsec-scale separations in complete samples of ``radio-quiet'' AGN \citep{Rosario11}. The search for binary black holes with parsec-scale separations requires the technique of very long baseline interferometry (VLBI). Here we briefly describe the VLBI observations of two double peaked emission line Seyfert galaxies residing in spiral hosts.

\section{Seyfert Galaxies with Double Peaked Emission Lines}
We identified six DPAGN in the KPNO Internal Spectroscopic Survey Red \citep[KISSR;][]{Wegner03} comprising overall of 72 Seyfert and Low-Ionization Nuclear Emission-Line Region (LINER) galaxies, through their SDSS\footnote{Sloan Digital Sky Survey} optical spectra. The fraction of DPAGN in the KISSR Seyfert+LINER sample is therefore $\approx$8\%. However only three of these six Seyfert galaxies were detected in the VLA FIRST\footnote{Faint Images of the Radio Sky at Twenty-Centimeters} and NVSS\footnote{The NRAO VLA Sky Survey} surveys, and were suitable targets for carrying out a phase-referenced VLBI study. 
Results from the VLBI study of KISSR\,1494 and KISSR\,1219 are presented below.

\noindent
{\bf KISSR\,1494:}
VLBI observations of the Seyfert galaxy, KISSR\,1494 ($z=0.0574$), revealed a single radio core of size $\sim7.5\times5$~milliarcsecond (=$8\times6$ parsec) with a flux density of $\sim650\,\mu$Jy at 1.6~GHz \citep{Kharb15}. There was no core detection at 5 GHz. The core brightness temperature ($T_b$) of $\sim1.4\times10^7$\,K was too high to be from a nuclear starburst which typically have $T_b<10^5$\,K, while the $1.6-5$ GHz radio spectral index of $-1.5\pm0.5$ was too steep to be from thermal free-free emission. These parameters are similar to those derived in ``normal'' Seyfert galaxies (ones with single peaked emission lines) that have parsec-scale radio outflows \citep[e.g.,][]{Kharb10,Kharb14}. The steep radio spectrum was also inconsistent with the KISSR\,1494 core being the unresolved base of a relativistic jet, which is expected to have a flat or inverted spectrum. We therefore concluded in \citet{Kharb15} that the radio emission was non-thermal synchrotron in origin that was likely coming from the base of an extended outflowing coronal wind. The NLR clouds were probably being pushed out by this broad nuclear outflow, giving rise to the double peaks in the optical spectrum.

\noindent
{\bf KISSR\,1219:}
VLBI observations of the second Seyfert galaxy, KISSR\,1219 ($z=0.0375$), took place in February 2015. A clear 100 milliarcsecond ($\sim$70 parsec) one-sided core-jet structure is revealed in our preliminary 1.6~GHz image (Kharb et al 2016, in preparation). The core but not the jet is detected at 5 GHz, indicating that the jet has a steep radio spectrum. Interestingly, this parsec-scale jet is pointing directly towards some lobe-like radio emission about 50 kiloparsec away, as seen in its VLA FIRST image (see Figure 2). If the parsec-scale and the kiloparsec-scale emission are connected, KISSR\,1219 will be the fourth only Seyfert galaxy with a large-scale jet ejected from a spiral galaxy \citep[see][]{Bagchi14}. New Expanded VLA observations of KISSR\,1219 are currently being analysed to search for this connection. Like in KISSR\,1494, there is an indication that the double peaked emission lines in KISSR\,1219 are a result of an AGN outflow impacting the emission line clouds.

\section{Summary}
Double peaked emission lines could indicate the presence of binary black holes, rotating BLR/NLR disks, or AGN outflows impacting emission-line clouds. Parsec-scale outflows or binary black holes with parsec-scale separations can be searched through the technique of VLBI. In this paper, we briefly describe results from phase-referenced VLBI observations at 1.6 and 5 GHz of two DPAGN from the KISSR Seyfert+LINER sample, residing in spiral host galaxies. 

A single slightly resolved radio core was detected at 1.6~GHz in the Seyfert galaxy KISSR\,1494. The high brightness temperature and the steep $1.6-5$ GHz spectral index were consistent with the core being the base of an extended outflowing synchrotron-emitting coronal wind, rather than the unresolved compact base of a radio jet \citep{Kharb15}. A clear $\sim$70 parsec radio jet is revealed, on the other hand, in the 1.6~GHz VLBI image of the Seyfert galaxy KISSR\,1219. 
The VLBI observations therefore do not reveal any signatures of binary black holes in the two Seyfert galaxies, but rather those of an AGN-related outflow. We conclude that an AGN outflow$-$emission line cloud interaction is the most likely cause of the splitting of emission line peaks in these Seyfert galaxies.

\noindent
{\bf Acknowledgements}
We thank the referee for helpful suggestions. PK would like to thank the RETCO-II meeting organisers and Dr. Indranil Chattopadhyay for this invited talk. ZP was supported by the International Space Science Institute in Bern. 

\bibliographystyle{apj}
\bibliography{PKharb}

\begin{figure}
\centering{
\includegraphics[width=12.5cm,trim=20 95 0 100]{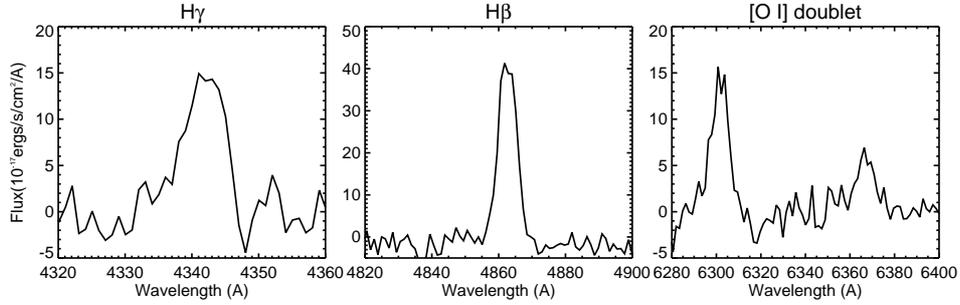}}
\caption{The double peaked emission line SDSS spectrum of the Seyfert galaxy KISSR\,1219. }
\end{figure}
\begin{figure}
\centering{
\includegraphics[width=9cm,trim=50 240 50 220]{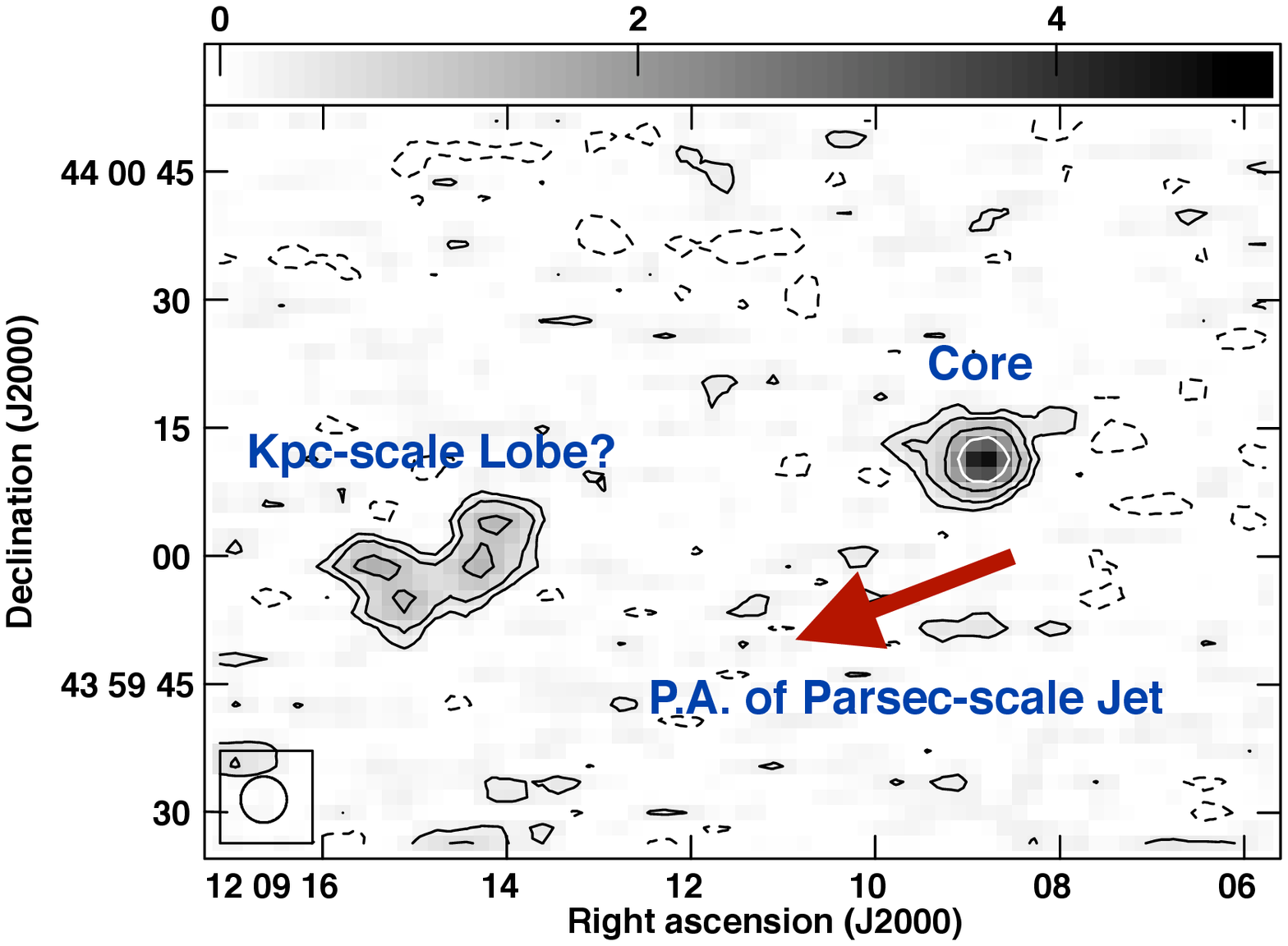}}
\caption{The VLA FIRST image at 1.4 GHz of KISSR\,1219. The VLBI jet in this source is at the same position angle (P.A.) as the lobe-like emission seen $\approx65^{\prime\prime}$ (=50~kiloparsec) away to the south east of the radio core. The P.A. of the VLBI jet and the position of the kiloparsec-scale emission are noted in the figure.}
\end{figure}

\label{lastpage}
\end{document}